\documentclass[x11names,12pt]{article}
\usepackage[paper=letterpaper,margin=.85in]{geometry}

\pdfoutput=1

\usepackage{graphicx}				
\usepackage{amsmath}
\usepackage{mathtools}
\usepackage{amssymb}
\usepackage{booktabs}
\usepackage{comment}
\usepackage{amsthm}
\numberwithin{equation}{section}

\usepackage{multirow}
\usepackage{overpic}
\usepackage{caption}
\usepackage{subcaption}
\usepackage{array}
\usepackage{rotating}

\usepackage[utf8]{inputenc}
\usepackage{lmodern}
\usepackage[T1]{fontenc} 
\usepackage{microtype} 

\usepackage{xcolor}
\definecolor{dark-green}{rgb}{0.1,0.4,0}
\definecolor{NiceBlue}{rgb}{0.30196,0.55294,0.57647}

\usepackage{cite}
\usepackage{hyperref}
\hypersetup{
      colorlinks=true,
      linkcolor=dark-green,
      citecolor=Red4,
      urlcolor=dark-green,
      linktoc=page
}
\newcommand{\bea}{\begin{eqnarray}}
\newcommand{\eea}{\end{eqnarray}}

\newcommand\afterTocSpace{\bigskip\medskip}
\newcommand\afterTocRuleSpace{\bigskip\bigskip}

\newcommand{\dd}{\mathrm{d}}
\newcommand{\brk}{\text{brk}}
\newcommand{\Ai}{\text{Ai}}

\newcommand\emailfootnote[1]{%
  \begingroup
  \renewcommand\thefootnote{}\footnote{#1}%
  \addtocounter{footnote}{-1}%
  \endgroup
}

\begin{document} 
\thispagestyle{empty}

\vspace*{1.5cm}
\begin{center}

{\bf {\LARGE Non-Perturbative Corrections to Charged \\ \vspace{7pt} Black Hole Evaporation}}

\begin{center}

\vspace{1cm}

{\bf Vyshnav Mohan}$^1$\emailfootnote{$^1$ \href{mailto:vyshnav.vijay.mohan@gmail.com}{vyshnav.vijay.mohan@gmail.com} }, \textbf{and} {\bf L\'arus Thorlacius}$^2$\emailfootnote{$^2$ \href{mailto:lth@hi.is}{lth@hi.is} }\\
 \bigskip \rm
  
\bigskip
\hspace{.05em}Science Institute,
University of Iceland \\Dunhaga 3, 107 Reykjav\'{i}k, Iceland.

\bigskip

\rm
  \end{center}

\vspace{2.5cm}
{\bf Abstract}
\end{center}
\begin{quotation}
\noindent

The recent work of Brown \textit{et al.} (arXiv:2411.03447) demonstrated that the low-temperature evaporation rate of a large near-extremal charged black hole is significantly reduced from semiclassical expectations. The quantum corrections responsible for the deviation come from Schwarzian modes of an emergent Jackiw-Teitelboim gravity description of the near-horizon geometry of the black hole. Using a one-parameter family of non-perturbative Airy completions, we extend these results to incorporate non-perturbative effects. At large parameter value, the non-perturbative evaporation rate is even smaller than the perturbative JT gravity results. The disparity becomes especially pronounced at very low energies, where the non-perturbative neutral Hawking flux is suppressed by a double exponential in the entropy of the black hole, effectively stopping its evaporation until the next charged particle is emitted via the Schwinger effect. We also explore an alternative family of Bessel completions for which the non-perturbative energy flux exceeds the perturbative JT gravity prediction.
\end{quotation}

\setcounter{page}{0}
\setcounter{tocdepth}{2}
\setcounter{footnote}{0}
\newpage

\parskip 0.1in
 
\setcounter{page}{2}

\setcounter{tocdepth}{1}
{\hypersetup{linkcolor=black}
\tableofcontents
}

\section{Introduction}
\label{introduction}
Jackiw-Teitelboim (JT) gravity \cite{Jackiw:1984je,Teitelboim:1983ux} and the associated Schwarzian dynamics serve as a robust framework for systematically incorporating quantum gravity corrections into perturbative QFT computations on a black hole background \cite{Penington:2019kki,Saad:2019pqd}. An interesting recent development is the computation of the low-temperature evaporation rate of a large near-extremal black hole \cite{Brown:2024ajk}, where it was shown that the semiclassical picture of Hawking radiation breaks down under the following conditions (in Planck units),
\bea
E_i \equiv M-Q\lesssim E_{\brk} \equiv \frac{1}{Q^3}\, . \label{defofebrk}
\eea 
The quantum corrections from Schwarzian modes of an emergent JT gravity description in the near-horizon region of the black hole become significant in this regime. When energies are below $E_{\brk}$, these modes are strongly coupled, and the density of states gets modified. In particular, the leading perturbative density of states $\rho_0(E)$ goes to zero when $E$ goes to zero, without producing a mass gap in the spectrum \cite{Iliesiu:2020qvm}.

The density of states dropping to zero at $E=0$ has important consequences for the computation of the black hole evaporation rate. When $E_i \ll E_\brk$, the Schwarzian spectrum constrains the emission of particles with energies greater than $E_i$. This sharply contrasts the semiclassical behaviour, where no such restriction exists. The resulting modifications to the evaporation rates were analyzed extensively in \cite{Brown:2024ajk}.

In this paper, we study the effect of non-perturbative corrections on these emission rates using non-perturbative 
completions of JT gravity based random matrix models as pioneered in \cite{Saad:2019lba} and further 
developed in \cite{Johnson:2019eik}. 
We start with a one-parameter family of Airy completions that were introduced in \cite{Johnson:2019eik}.
The parameter brings a new energy scale, $\mu$, into the problem which affects the rate of evaporation at 
low energies. These Airy completions suffer from a non-perturbative instability which can, however, be avoided via
more intricate matrix model constructions as described in \cite{Johnson:2019eik}. We mainly focus on the Airy
case, due to its relative simplicity, and address the non-perturbative instability by simply truncating the Airy density
of states at $E_i=0$. Towards the end of the paper we also consider a family of Bessel completions, that are 
argued to be stable in \cite{Johnson:2019eik}. 

\begin{figure}
\centering
\includegraphics[width=0.8\linewidth]{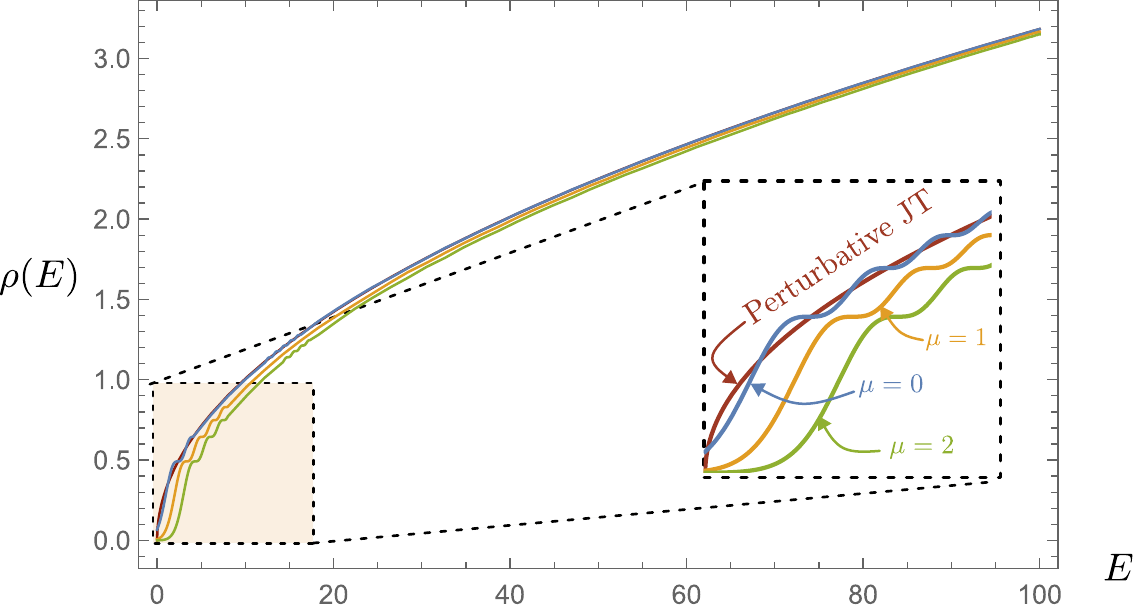}
\caption{The plot of the density of states our non-perturbative Airy completions of JT gravity, with the leading perturbative result shown in red. These completions, labelled by a parameter $\mu$, exhibit an exponential decay at low energies.} \label{airyplot}
\end{figure}

In figure \ref{airyplot}, we have plotted the density of states of a few non-perturbative Airy completions. The plots exhibit two new features that are absent in the perturbative density of states. The plots contain wiggles that die off at large energies, a direct consequence of the discreteness of the underlying spectrum. The other significant feature is an exponential suppression of the density of states at low energies, controlled by the emergent energy scale $\mu$. The combination of the wiggles and the low energy density of states alters the evaporation rate compared to the perturbative result of \cite{Brown:2024ajk}.\footnote{In this paper, we will reserve the term `perturbative' for leading order disc computations in JT gravity. We will use the term `semiclassical' to refer to perturbative QFT calculations on the black hole background.}

Consider a black hole with integer spin. The low-temperature evaporation will be dominated by the emission of di-photons \cite{Brown:2024ajk}. When $\mu \gg e^{-2S_0/3}E_\brk$, we find the non-perturbative evaporation rates to be:
\bea
\frac{\dd M}{\dd t} \sim \begin{dcases}-(M-Q)^5, & M-Q \gg E_{\mathrm{brk} \,,} \\ -(M-Q)^{\frac{19}{2}}, & \mu \ll M-Q \ll E_{\mathrm{brk} \,,} \\ -\exp\left(-\frac{4e^{S_0}\mu^{3/2}}{E_{\brk}^{3/2}}\right) (M-Q)^{9}, & M-Q \ll \mu \,. \end{dcases}
\eea 
When $M-Q \gg \mu$, we recover the perturbative and semiclassical results of \cite{Brown:2024ajk}. When $M-Q \ll \mu$, the evaporation rate is suppressed by a double exponential in the entropy of the black hole. This is because the exponentially decaying tail of the density of states essentially cuts off the emission of all particles (see figure \ref{airyplot}). Therefore, the black hole has more or less stopped evaporating.

The black hole also has a probability of emitting a positron via the Schwinger effect. Upon the emission of a positron, the black hole is pushed away from extremality. Positron emission can, in principle, compete with our non-perturbative corrections, as it may push the black hole out of extremality before it reaches the energy scale necessary to observe the new effects. However, since positron emission is important only at energy scales of the order of $e^{-O\left(\sqrt{S_0}\right)}$ \cite{Brown:2024ajk}, our non-perturbative effects at $M-Q \gg \mu$ become noticeable well before the black hole emits a positron and is driven away from extremality.

When $\mu \ll e^{-2S_0/3}E_\brk$, non-perturbative corrections become important at $E_i \sim e^{-2S_0/3}E_\brk$. We find that non-perturbative emission in fact exceeds the perturbative result of \cite{Brown:2024ajk} at such low energies but, unfortunately, positron emission is also relevant at these energies and is likely to move the black hole away from extremality before any non-perturbative enhancement of the emission rate can be observed.

One might have expected the emergent non-perturbative energy scale $\mu$ to always be smaller than $O(e^{-S_0})$ in suitable units of energy. However, we note that the Airy completions presented here should be viewed as simple avatars of more realistic models of non-perturbative JT gravity \cite{Johnson:2019eik}. Therefore, one should ask whether non-perturbative completions of JT gravity exist in which the analog of the parameter $\mu$ is not exponentially suppressed. We find the answer to be affirmative, as there are solutions to the master string equation where a specific half-integer parameter, $\Gamma$, is turned on (see figure 9 of \cite{Johnson:2020heh} and the discussion around it). The non-perturbative sector of these models does not contain D-branes, and as a result, non-perturbative corrections are not exponentially suppressed in the inverse of the string coupling constant \cite{Johnson:2006ux}. This justifies considering larger values of $\mu$.

We can also perform a Bessel completion of the density of states. In this case, the eigenvalues of the corresponding random matrix `pile' up at $E=0$, and this is expected to capture the low energy physics of stable non-perturbative completions of JT gravity and its supergravity counterparts \cite{Stanford:2019vob,Johnson:2019eik,Johnson:2020heh}. In Section~\ref{sec:bessel} below, we find that the pile-up of eigenvalues leads to an energy flux at low energies that significantly exceeds the perturbative result.

Let us end this introduction with the following remark. As proposed in \cite{Brown:2024ajk}, the energy flux from a large near-extremal charged black hole can in principle be measured using a simple spectrograph if a large near-extremal astrophysical black hole is ever observed. The main result of the present paper is that such observations would not only demonstrate a novel breakdown of semi-classical gravity, as was emphasised in \cite{Brown:2024ajk}, but offer a direct window onto non-perturbative quantum effects. They could even allow observers to distinguish between different non-perturbative completions of JT gravity. Given that the energy scale $\mu$ is sufficiently large in some of these completions, the distinction could be made without having to wait for exponentially long time scales. 

\section{Non-perturbative JT Gravity}
The spherical reduction of Einstein-Maxwell theory in the near-horizon region of a near-extremal Reissner-Nordstr\"{o}m black hole results in JT gravity \cite{Navarro-Salas:1999zer,Nayak:2018qej}, whose action is given by
\bea
I = S_0 +\frac{1}{2} \int d^2 x \sqrt{g} \Phi\left(R+\frac{2}{\ell^2}\right)+\Phi_{b} \int_{\partial_M} d u \sqrt{h}\left(K-\frac{1}{\ell}\right)\, .
\eea 
Here $S_0$ is the extremal entropy of the black hole, and $\ell$ is the AdS$_2$ length scale. The extremal entropy and the AdS$_2$ length scale are related to the extremal charge $Q$ through the relations
\bea
S_0 = \pi Q^2\,, \qquad \ell =Q \,.
\eea 
When the energy above extremality is below $E_{\brk}$, Schwarzian dynamics of the emergent JT gravity description becomes important. The Schwarzian description offers analytic control and the theory can be non-perturbatively completed using the connection between JT gravity and matrix models. The two-dimensional dilaton theory is dual to a double-scaled Hermitian matrix model \cite{Saad:2019lba}, whose leading perturbative density of states is given by
\bea
\rho_0(E, Q)=\frac{e^{S_0(Q)}}{2 \pi^2 E_{\text {brk }}} \sinh \left(2 \pi \sqrt{2} \sqrt{\frac{E}{E_{\text {brk }}}}\right) \Theta(E)\, . \label{JTdos}
\eea 
In \cite{Brown:2024ajk}, only the perturbative JT gravity corrections were considered. We extend this analysis by including  non-perturbative corrections.

There is no unique non-perturbative completion of JT gravity (see \cite{Saad:2019lba,Johnson:2019eik,Collier:2023cyw} for some examples). We will use an Airy completion, that is central to the work of \cite{Saad:2019lba}, to model the non-perturbative completions in \cite{Johnson:2019eik}. The exact Airy density of states is given by 
\bea
\rho_\text{Ai}(E) = h^{-\frac{2}{3}}\left[\Ai^{\prime}(\zeta)^2-\zeta\Ai(\zeta)^2\right] \Theta(E)\, , \label{exactairydos}
\eea 
where $\zeta = -h^{-\frac{2}{3}} \left(E-\mu\right)$. The $\Theta$ function in \eqref{exactairydos} simulates the natural truncation at $E=0$ of the density of states in \cite{Johnson:2019eik}, as was explained in section~\ref{introduction}. The parameter $\mu$ is chosen to be positive to model key features of the non-perturbative completions discussed in \cite{Johnson:2019eik}. When $E \gg \mu$, the function asymptotes to (see figure \ref{airyplot})
\bea
\rho_{\Ai}(E)\simeq \frac{\sqrt{E}}{\pi h}\, . \label{perturbativeairydos}
\eea 
When $E \ll E_{\brk}$, the perturbative JT gravity density of states also approaches \eqref{perturbativeairydos}, with 
\bea
h = \frac{E_{\brk}^{\frac{3}{2}}}{e^{S_0}\sqrt{2}}\, . \label{hdef}
\eea 
Therefore, we can use the exact Airy density of states to model non-perturbative JT gravity at low energies. To ensure that the non-perturbative completion eventually asymptotes to \eqref{perturbativeairydos}, we will assume that
\bea
\mu \ll E_\brk\, .
\eea 
\section{Massless Scalar Emission}
\begin{figure}[t!]
  \centering
  \begin{subfigure}[b]{0.49\textwidth}
  \centering
  \includegraphics[width=\textwidth]{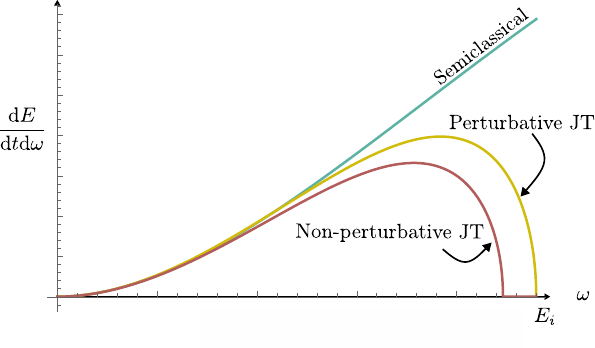}
  \caption{}
  \label{masslesscompplot}
  \end{subfigure}
  \begin{subfigure}[b]{0.49\textwidth} 
  \centering 
  \includegraphics[width=\textwidth]{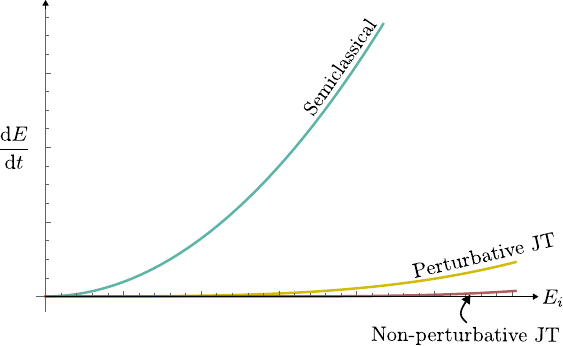}
  \caption{}
  \label{masslessfluxplots}
  \end{subfigure}
  \caption{Plots of energy flux from massless scalar particle emission when $E_i \sim \mu$. \textbf{(a)} The non-perturbative density of states has an exponentially decaying tail that stops the emission of particles of energy in the vicinity of $E_i$. \textbf{(b)} The total non-perturbative energy flux is smaller than that of the perturbative and semiclassical results.} 
  \label{masslessplots}
  \end{figure}
In this section, we will study the emission of massless scalar particles as it provides a useful testing ground for working out non-perturbative corrections. We will not repeat the computations in \cite{Brown:2024ajk} but directly quote relevant results here. The microcanonical massless scalar particle energy flux per unit time is given by \cite{Brown:2024ajk}\footnote{Schwarzian corrections to the energy flux were also independently studied in \cite{Bai:2023hpd}.}:
\bea
\frac{\dd E}{\dd t}=4\pi E_{\brk}e^{-S_0}\int_0^{E_i} d \omega \omega\left(r_{+} \omega\right)^2 \frac{\rho\left(E_i-\omega\right)}{\cosh \left(2 \pi \sqrt{2 E_{\mathrm{brk}}^{-1} E_i}\right)-\cosh \left(2 \pi \sqrt{2 E_{\mathrm{brk}}^{-1}\left(E_i-\omega\right)}\right)}\, . \label{masslessfluxexp}
\eea
Here, $E_i$ is the energy of the black hole, and $r_+$ is the radius of the outer horizon. In \cite{Brown:2024ajk}, the density of states was chosen to be the leading disk density of states \eqref{JTdos}. However, non-perturbative corrections kick in when $E_i$ goes to zero, requiring us to modify the density of states in \eqref{masslessfluxexp}. The rest of the integral remains the same because the matrix elements of operators do not change when upon the addition of non-perturbative corrections (see \cite{Saad:2019pqd} and \cite{Iliesiu:2021ari} for examples). Let us first look at the case where $\mu \gg h^{2/3}$, where $h$ is defined in \eqref{hdef}. We will return to the other case at the tail end of this section. Now, let us look at the energy flux at different energy scales.

\textbf{Semiclassical limit} ($E_i/E_{\brk} \to \infty$) : In this limit, the non-perturbative corrections are subleading, and we can use the perturbative density of states \eqref{JTdos} in the integral \eqref{masslessfluxexp}. Approximating hyperbolic functions by exponentials, we reproduce the result in \cite{Brown:2024ajk}:
\bea
\hspace{4.25cm}\frac{\dd E}{\dd t}=\frac{1}{30 \pi} r_{+}^2 E_{\mathrm{brk} }^2 E_i^2\, .\hspace{4.75cm} \textcolor{teal}{\text{Semiclassical}} \label{semiclassicalmassless}
\eea

\textbf{Quantum limit} ($E_i, \omega \ll E_{\brk} $): Using the Airy density of states in the integral \eqref{masslessfluxexp}, we get:
\bea
\frac{\dd E}{\dd t} = \frac{E_{\brk}^2}{\pi e^{S_0}}\int_0^{E_i} (r_+ \omega)^2 \rho_{\Ai}(E_i-\omega)\, . \label{airymasslessintegrand}
\eea 
Before explicitly evaluating the integral, let us look at the following cases separately.

\underline{Case 1} : $\mu \ll E_i, \omega \ll E_{\brk}\,$.
The non-perturbative corrections are subleading, and the Airy density of states reduces to \eqref{perturbativeairydos}. This gives us the perturbative JT gravity result in \cite{Brown:2024ajk}: 
\bea
\hspace{3cm}\frac{\dd E}{\dd t} = \frac{16\sqrt{2E_{\brk}}r_+^2}{105 \pi^2} \ E_i^{\frac{7}{2}}\, . \hspace{5.75cm} \textcolor{teal}{\text{Perturbative}} \label{perturbativemassless}
\eea

\underline{Case 2} : $ E_i, \omega \lesssim \mu$.
In this regime, the non-perturbative effects become important. 
Integrating \eqref{airymasslessintegrand}, we get 
{
\begin{align}
\frac{\dd E}{\dd t} = \frac{\sqrt[3]{4}r_+^2 e^{S_0/3}}{105\pi} \Bigg[ &\left(21 h^2 \mu - 70 E_i^2 \mu^2 - 16 \mu^4 - 
7 E_i (3 h^2 - 8 \mu^3)\right) \left(\text{Ai}\left(\frac{\mu}{h^{
2/3}}\right)\right)^2 \nonumber\\
&+ (E_i - \mu) \left(21 h^2 + 16 (E_i - \mu)^3\right) \left(\text{Ai}\left(\frac{-E_i + \mu}{h^{2/3}}\right)\right)^2 \nonumber\\ 
&-h^{4/3} (-35 E_i^2 + 28 E_i \mu - 8 \mu^2) \text{Ai}\left(\frac{\mu}{h^{
 2/3}}\right) \text{Ai}^{\prime}\left(\frac{\mu}{h^{
  2/3}}\right) \nonumber \\
&-8 h^{4/3} (E_i - \mu)^2 \text{Ai}\left(\frac{-E_i + \mu}{h^{2/3}}\right) \text{Ai}^{\prime}\left(\frac{-E_i + \mu}{h^{2/3}}\right)\hspace{1.5cm} \textcolor{teal}{\text{Non-perturbative}} \nonumber\\ 
&- h^{2/3} \left(15 h^2 - 70 E_i^2 \mu + 56 E_i \mu^2 - 
16 \mu^3 \right) \left(\text{Ai}^{\prime}\left(\frac{\mu}{h^{2/3}}\right)\right)^2 \nonumber\\ 
 &+h^{2/3} \left(15 h^2 + 16 (E_i - \mu)^3\right) \left(\text{Ai}^{\prime}\left(\frac{-E_i + \mu}{h^{2/3}}\right)\right)^2
\Bigg] \, . \label{nonperturbativemassless}
\end{align}}

\noindent Plotting \eqref{semiclassicalmassless}, \eqref{perturbativemassless} and \eqref{nonperturbativemassless} in figure \ref{masslessfluxplots}, we find the non-perturbative flux to be smaller than both the perturbative and semiclassical answers. A straightforward way to see this is to look at the integrand of \eqref{airymasslessintegrand}. The Airy density of states has an exponentially suppressed tail near the lower edge of the spectrum. This prevents the emission of particles with energy in the neighbourhood of $E_i$, lowering the flux (see figure \ref{masslesscompplot}).

\noindent When $E_i \ll \mu$, something very interesting happens. Expanding \eqref{nonperturbativemassless} in powers of $E_i/E_{\brk}$, we find that
\bea
\begin{aligned}
\frac{\dd E}{\dd t} &= e^{\frac{-S_0}{3}}\left(\frac{\sqrt[3]{2}r_+^2}{3 \pi}\right) \left( E_{\brk} \text{Ai}'\left(\frac{\sqrt[3]{2} e^{\frac{2 S_0}{3}} \mu }{E_{\brk}}\right)^2-\sqrt[3]{2} \mu e^{\frac{2 S_0}{3}} \text{Ai}\left(\frac{\sqrt[3]{2} e^{\frac{2 S_0}{3}} \mu }{E_\brk}\right)^2\right) E_i^3 \\
&\sim \exp\left(-\frac{4 \sqrt{2}e^{S_0}\mu^{3/2}}{3E_{\brk}^{3/2}}\right) \frac{E_{\brk}^2 r_+^2}{24 \pi^2 \mu } E_i^3\, .
\end{aligned}
\eea
Therefore, the flux rate is suppressed by a double exponential in the entropy of the black hole. As before, the suppression comes from the exponential tail of the density of states, which suppresses the emission of low-energy particles. In the next section, we will see that the same phenomenon also suppresses the photon evaporation rate of the black hole in this parameter range.

Let us return to the case where $\mu \lesssim h^{2/3}$. The non-perturbative effects kick in only when $E_i \lesssim h^{2/3}$. The non-perturbative flux, given in \eqref{nonperturbativemassless}, is larger than the perturbative answer. We can see this again by looking at the integrand of \eqref{airymasslessintegrand}. A helpful example to remember is the $\mu=0$ plot in figure \ref{airyplot}. The non-perturbative density of states remains finite in the $E\rightarrow 0$ limit while the perturbative one goes as $\sqrt{E}$, and as a result, the non-perturbative integral will be larger.

\section{Photon Emission}
When $E_i\leq E_\brk$, the dominant decay channel of the black hole is through the emission of di-photons. The exact quantum corrected energy flux can be found in equation (3.73) of \cite{Brown:2024ajk}. The Clebsch-Gordan coefficients come from the matrix elements of the corresponding JT gravity operator insertions and, as a result, will not change upon non-perturbative completion. Moreover, the intermediate energies in the expression can be assumed to be above the breakdown scale when $E_i \ll E_{\brk}\,$. Therefore, we can borrow the processed result of \cite{Brown:2024ajk}
\bea
\frac{\dd E}{\dd t} =\left(1.4 \times 10^{-1}\right) \frac{8}{3 \pi^3} E_{\text {brk}}^{10} r_{+}^{16} \int_0^{E_i} d E_f e^{-S_0} \rho\left(E_f\right) \int_0^{E_i-E_f} d \omega \omega^3\left(E_i-E_f-\omega\right)^3\left(E_i-E_f\right)\, , \nonumber\\\label{diphotonflux}
\eea
and use the Airy density of states to get non-perturbative corrections. As in the previous section, it is convenient to examine the expression separately at different energy scales. We reproduce the perturbative JT gravity result
\begin{figure}
 \centering
 \includegraphics[width=0.65\linewidth]{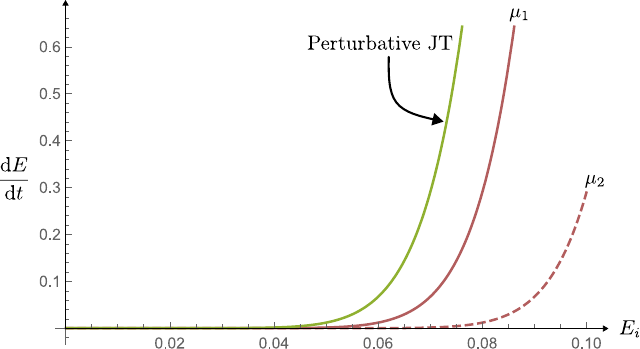}
 \caption{The plot shows the di-photon emission flux when $E_i \lesssim \mu$. The perturbative JT result is shown in green. The red curves correspond to non-perturbative flux rates for two different values of $\mu$. Here, $\mu_2>\mu_1\gg h^{2/3}$. }
 \label{diphotonplot}
 \end{figure}
\begin{align}
\frac{\dd E}{\dd t} \sim r_{+}^{16} E_\brk^{\frac{17}{2}} E_i^{\frac{19}{2}}\, , \quad \quad \text{when} \ \mu \ll E_i \ll E_\brk\,.\label{perturbativediphoton}
\end{align}

When $E_i \lesssim \mu$, we can numerically integrate \eqref{diphotonflux}. Let us first look at the case where $\mu \gg h^{2/3}$. We see that the non-perturbative flux is less than the perturbative result. Moreover, the flux decreases as the value of $\mu$ increases (see figure \ref{diphotonplot}). The reduction in flux can be traced back to the exponential suppression of the density of states at energies lower than the scale set by $\mu$, as in the case of massless scalar particles.

\noindent When $E_f \ll \mu$, there is an even more dramatic dip in the energy flux. To see this, let us expand the integrand in powers of $E_f/E_\brk\,$:
\bea
\begin{aligned}
\frac{\dd E}{\dd t} = &\, \left( 1.1\times 10^{-4} \right) \, E_{\brk}^8 \, r_+^{16} \, e^{S/3} \, E_i^9 \Bigg( 
E_{\brk} \, e^{-\frac{2S_0}{3}} \, \text{Ai}'\left(\frac{\sqrt[3]{2} e^{\frac{2S_0}{3}} \, \mu}{E_{\brk}}\right)^2 \\
&\qquad - \sqrt[3]{2} \, \mu \, \text{Ai}\left(\frac{\sqrt[3]{2} e^{\frac{2S_0}{3}} \, \mu}{E_{\brk}}\right)^2 \Bigg) \, .
\end{aligned}\label{diphotonsmallenergyeq}
\eea 
Expanding the above expression for large values of $\frac{\mu}{h^{2/3}}$, we find that
\bea
\frac{\dd E}{\dd t} \sim \exp\left(-\frac{4 \sqrt{2}e^{S_0}\mu^{3/2}}{3E_{\brk}^{3/2}}\right) \frac{E_\brk^{10} r_+^{16}}{\mu}E_i^9\, .\label{diphotonsuppression}
\eea 
The flux is suppressed by a double exponential in the entropy of the black hole. Since di-photon emission dominates over graviphoton emissions at low energies, the change in the mass of a bosonic black hole comes mostly from the di-photon emissions \cite{Brown:2024ajk}. As a result, the suppression in \eqref{diphotonsuppression} has important consequences. Ignoring positron emissions, we find that (in Planck units)
\bea
\frac{\dd M}{\dd t} \sim \begin{dcases}-(M-Q)^5, & M-Q \gg E_{\mathrm{brk} \,,} \\ -(M-Q)^{\frac{19}{2}}, & \mu \ll M-Q \ll E_{\mathrm{brk} \,,} \\ -\exp\left(-\frac{4e^{S_0}\mu^{3/2}}{E_{\brk}^{3/2}}\right) (M-Q)^{9}, & M-Q \ll \mu\,. \end{dcases} \label{bosonicevaporationrate}
\eea 
When $M-Q \gg E_\brk$, the non-perturbative effects are subleading, and the evaporation rate reduces to the single-photon computation in \cite{Brown:2024ajk}. We have displayed their results in the first line of \eqref{bosonicevaporationrate}. When the energy above extremality drops below $\mu$, the black hole has effectively stopped evaporating!

Now let us look at the case where $\mu \ll h^{2/3}$. We can expand \eqref{diphotonsmallenergyeq} for small values of $\frac{\mu}{h^{2/3}}$ to find
\bea
\frac{\dd E}{\dd t} \sim e^{-\frac{S}{3}} E_\brk^9 E_i^9 r_+^{16} \, .
\eea 
When $E_f \ll \mu$, $E_i/E_\brk \ll e^{-2S/3}$. Therefore, the non-perturbative flux is larger than the perturbative answer \eqref{perturbativediphoton}. As in the massless case, we can read this off from the plot of the density of states of the $\mu=0$ completion in figure \ref{airyplot}. Consequently, the evaporation due to di-photon emission increases upon non-perturbative completion. As discussed in the Introduction, the emission of positron will undermine this conclusion as the black hole is likely to be pushed away from of extremality before these corrections kick in. 

A similar analysis can be performed for the emission of particles with spin. Since the additional $SU(2)$ modes decouple from the JT gravity modes \cite{Brown:2024ajk}, all we have to do is replace the low energy JT density of states with the Airy density of states. Since the inferences one draws are the same as those for massless scalar and photon emissions, we will not go through the calculations here. 

\section{When Eigenvalues Pile Up: The Bessel Completion}\label{sec:bessel}
\begin{figure}
\centering
\includegraphics[width=1.02\linewidth]{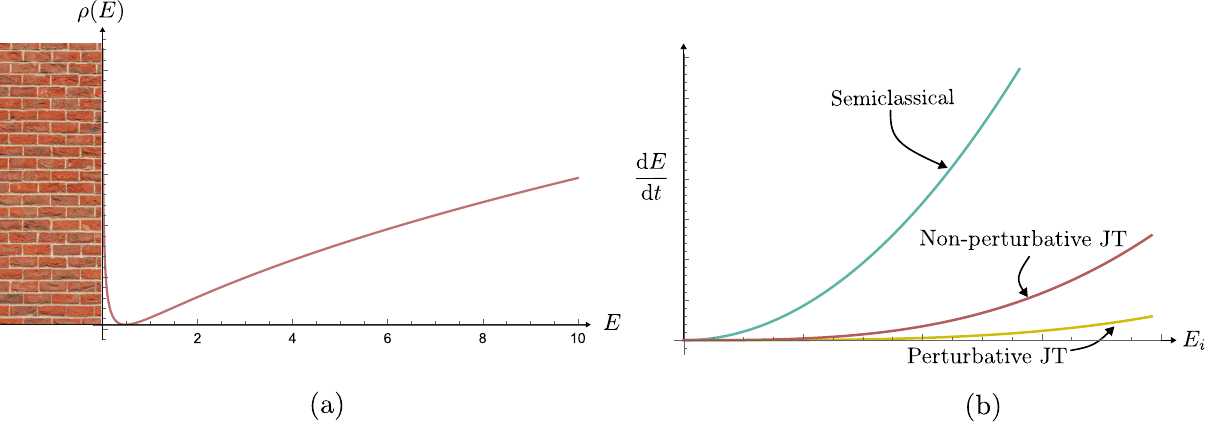}
\caption{\textbf{(a)}. The plot of the density of states of a Bessel completion. The eigenvalues pile up against a ``wall'', and the value of $\rho(E)$ at $E=0$ depends on the parameter $\mu$. \textbf{(b)}. The non-perturbative flux lies between the semiclassical and perturbative plots.}
\label{besselplot}
\end{figure}
Another curious non-perturbative completion of JT gravity is where the eigenvalues pile up at $E=0$ (see figure \ref{besselplot}). Non-perturbative completions of this sort were found in \cite{Johnson:2019eik} and have the favourable behaviour that they do not have any incursions into the $E<0$ forbidden region. The incursion is stopped by an infinite ``wall,'' which emerges from integrating out angular variables in complex matrix models \cite{Morris:1990bw,Dalley:1991qg}. We model this behaviour by working with the following density of states:
\bea
\rho(E) = \rho_0(E)+ \frac{E_\brk}{2\pi h \sqrt{E+\mu}}\, ,
\eea 
where $h$ is given by \eqref{hdef} and $\rho_0(E)$ is the perturbative JT gravity density of states. The second term in the above expression asymptotes to the low-energy density of state in the Bessel models of \cite{Johnson:2019eik}, which in turn are expected to be a good description of the low energy behaviour of JT supergravity \cite{Stanford:2019vob,Johnson:2020mwi}. The parameter $\mu$ controls the value of the density of states at $E=0$.

The piling up of eigenvalues results in an increased flux at low energies. To see this, let us revisit the massless scalar particle emission computation. When $E_i \ll \mu$, we can integrate the flux to obtain
\bea
\frac{\dd E}{\dd t} \simeq \frac{E_\brk^{3/2} r_+^2}{3 \sqrt{2} \pi ^2 \sqrt{\mu}} E_i^3\,.
\eea
The non-perturbative answer lies between the semiclassical \eqref{semiclassicalmassless} and perturbative expressions \eqref{perturbativemassless} (see figure \ref{besselplot}). 

In  \cite{Brown:2024ajk} it was shown that at late times the flux of neutral radiation from a large charged black hole will be significantly reduced compared to semiclassical expectations. Overall, our results indicate that any measurement sensitive enough to pick up the perturbative JT gravity signal predicted by \cite{Brown:2024ajk} could in principle also differentiate among various suggested non-perturbative completions and place constraints on their parameter values.

\section*{Acknowledgements} 
We would like to thank Matthias Harksen for useful discussions. This work was supported in part by the Icelandic Research Fund under grant 228952-053. VM is supported by a doctoral grant from The University of Iceland Science Park. 

\bibliographystyle{JHEP}
\bibliography{refs}

\providecommand{\href}[2]{#2}\begingroup\raggedright\begin{thebibliography}{10}

\bibitem{Jackiw:1984je}
R.~Jackiw, \emph{{Lower Dimensional Gravity}},
  \href{https://doi.org/10.1016/0550-3213(85)90448-1}{\emph{Nucl. Phys. B}
  {\bfseries 252} (1985) 343}.

\bibitem{Teitelboim:1983ux}
C.~Teitelboim, \emph{{Gravitation and Hamiltonian Structure in Two Space-Time
  Dimensions}}, \href{https://doi.org/10.1016/0370-2693(83)90012-6}{\emph{Phys.
  Lett. B} {\bfseries 126} (1983) 41}.

\bibitem{Penington:2019kki}
G.~Penington, S.H.~Shenker, D.~Stanford and Z.~Yang, \emph{{Replica wormholes
  and the black hole interior}},
  \href{https://doi.org/10.1007/JHEP03(2022)205}{\emph{JHEP} {\bfseries 03}
  (2022) 205} [\href{https://arxiv.org/abs/1911.11977}{{\ttfamily
  1911.11977}}].

\bibitem{Saad:2019pqd}
P.~Saad, \emph{{Late Time Correlation Functions, Baby Universes, and ETH in JT
  Gravity}},  \href{https://arxiv.org/abs/1910.10311}{{\ttfamily 1910.10311}}.

\bibitem{Brown:2024ajk}
A.R.~Brown, L.V.~Iliesiu, G.~Penington and M.~Usatyuk, \emph{{The evaporation
  of charged black holes}},  \href{https://arxiv.org/abs/2411.03447}{{\ttfamily
  2411.03447}}.

\bibitem{Iliesiu:2020qvm}
L.V.~Iliesiu and G.J.~Turiaci, \emph{{The statistical mechanics of
  near-extremal black holes}},
  \href{https://doi.org/10.1007/JHEP05(2021)145}{\emph{JHEP} {\bfseries 05}
  (2021) 145} [\href{https://arxiv.org/abs/2003.02860}{{\ttfamily
  2003.02860}}].

\bibitem{Saad:2019lba}
P.~Saad, S.H.~Shenker and D.~Stanford, \emph{{JT gravity as a matrix
  integral}},  \href{https://arxiv.org/abs/1903.11115}{{\ttfamily 1903.11115}}.

\bibitem{Johnson:2019eik}
C.V.~Johnson, \emph{{Nonperturbative Jackiw-Teitelboim gravity}},
  \href{https://doi.org/10.1103/PhysRevD.101.106023}{\emph{Phys. Rev. D}
  {\bfseries 101} (2020) 106023}
  [\href{https://arxiv.org/abs/1912.03637}{{\ttfamily 1912.03637}}].

\bibitem{Johnson:2020heh}
C.V.~Johnson, \emph{{Jackiw-Teitelboim supergravity, minimal strings, and
  matrix models}},
  \href{https://doi.org/10.1103/PhysRevD.103.046012}{\emph{Phys. Rev. D}
  {\bfseries 103} (2021) 046012}
  [\href{https://arxiv.org/abs/2005.01893}{{\ttfamily 2005.01893}}].

\bibitem{Johnson:2006ux}
C.V.~Johnson, \emph{{String theory without branes}},
  \href{https://arxiv.org/abs/hep-th/0610223}{{\ttfamily hep-th/0610223}}.

\bibitem{Stanford:2019vob}
D.~Stanford and E.~Witten, \emph{{JT gravity and the ensembles of random matrix
  theory}}, \href{https://doi.org/10.4310/ATMP.2020.v24.n6.a4}{\emph{Adv.
  Theor. Math. Phys.} {\bfseries 24} (2020) 1475}
  [\href{https://arxiv.org/abs/1907.03363}{{\ttfamily 1907.03363}}].

\bibitem{Navarro-Salas:1999zer}
J.~Navarro-Salas and P.~Navarro, \emph{{AdS(2) / CFT(1) correspondence and near
  extremal black hole entropy}},
  \href{https://doi.org/10.1016/S0550-3213(00)00165-6}{\emph{Nucl. Phys. B}
  {\bfseries 579} (2000) 250}
  [\href{https://arxiv.org/abs/hep-th/9910076}{{\ttfamily hep-th/9910076}}].

\bibitem{Nayak:2018qej}
P.~Nayak, A.~Shukla, R.M.~Soni, S.P.~Trivedi and V.~Vishal, \emph{{On the
  Dynamics of Near-Extremal Black Holes}},
  \href{https://doi.org/10.1007/JHEP09(2018)048}{\emph{JHEP} {\bfseries 09}
  (2018) 048} [\href{https://arxiv.org/abs/1802.09547}{{\ttfamily
  1802.09547}}].

\bibitem{Collier:2023cyw}
S.~Collier, L.~Eberhardt, B.~M\"uhlmann and V.A.~Rodriguez, \emph{{The Virasoro
  minimal string}},
  \href{https://doi.org/10.21468/SciPostPhys.16.2.057}{\emph{SciPost Phys.}
  {\bfseries 16} (2024) 057}
  [\href{https://arxiv.org/abs/2309.10846}{{\ttfamily 2309.10846}}].

\bibitem{Bai:2023hpd}
Y.~Bai and M.~Korwar, \emph{{Near-extremal charged black holes: greybody
  factors and evolution}},
  \href{https://doi.org/10.1007/JHEP03(2023)151}{\emph{JHEP} {\bfseries 03}
  (2023) 151} [\href{https://arxiv.org/abs/2301.07739}{{\ttfamily
  2301.07739}}].

\bibitem{Iliesiu:2021ari}
L.V.~Iliesiu, M.~Mezei and G.~S\'arosi, \emph{{The volume of the black hole
  interior at late times}},
  \href{https://doi.org/10.1007/JHEP07(2022)073}{\emph{JHEP} {\bfseries 07}
  (2022) 073} [\href{https://arxiv.org/abs/2107.06286}{{\ttfamily
  2107.06286}}].

\bibitem{Morris:1990bw}
T.R.~Morris, \emph{{2-D quantum gravity, multicritical matter and complex
  matrices}}, .

\bibitem{Dalley:1991qg}
S.~Dalley, C.V.~Johnson and T.R.~Morris, \emph{{Multicritical complex matrix
  models and nonperturbative 2-D quantum gravity}},
  \href{https://doi.org/10.1016/0550-3213(92)90217-Y}{\emph{Nucl. Phys. B}
  {\bfseries 368} (1992) 625}.

\bibitem{Johnson:2020mwi}
C.V.~Johnson, \emph{{Low Energy Thermodynamics of JT Gravity and
  Supergravity}},  \href{https://arxiv.org/abs/2008.13120}{{\ttfamily
  2008.13120}}.

\end{thebibliography}\endgroup

\end{document}